\documentclass[11pt,a4paper,twoside,groupcitations]{article}
\usepackage[]{latexsym,amsmath,amssymb}
\usepackage[]{epsfig}
\usepackage{color}
\usepackage{braket}
\usepackage{graphicx}
\usepackage{cite}
\usepackage{stackrel}
\usepackage{amsfonts, bbold, mathtools, physics, cancel, scalerel, stackengine, wasysym, subcaption}

\usepackage{booktabs,tabularx,array,amsmath,amssymb}
\newcolumntype{Y}{>{\raggedright\arraybackslash}X}
\newcommand{\delg}{\delta_g}   
\DeclareMathOperator{\Sym}{Sym^2}

\newcommand{\be}{\begin{eqnarray}}
\newcommand{\ee}{\end{eqnarray}}
\def\beq{\begin{equation}}
\def\eeq{\end{equation}}

\renewcommand{\d}{\mbox{${\rm d}$}}

\newcommand{\Det}{\text{Det}\,}

\title{\bf The effective field theory of the gravitational functional measure}
\author{
Henrique Knopki$^a$\thanks{E-mail: knopki@fisica.ufpr.br},
$\ $
Iber\^e Kuntz$^{a,b}$\thanks{E-mail: kuntz@ufpr.br}
\\
\\
$^a${\em Departamento de F\'isica, Universidade Federal do Paran\'a}
\\
{\em Rua Cel. Francisco Heráclito dos Santos, 100, 81531-980, Curitiba -- PR, Brazil}
\\~\\
$^b${\em SISSA, via Bonomea 265, 34136, Trieste, Italy}
}
\date{}
\begin{document}
\maketitle
\begin{abstract}
The gravitational path integral measure has been the subject of an increasing interest lately, and no conclusive answer yet exists for its correct form. In this paper, we adopt effective field theory techniques to shed light on this issue. We build the configuration-space metric as an energy expansion, including all possible terms that satisfy the underlying symmetries, and use it to define a Riemannian measure. We study the running of the free parameters that show up in this expansion at leading order, which corresponds to the DeWitt metric with parameter $\lambda$. We show that a flat configuration space is excluded on unitarity grounds. The renormalization group contains one UV fixed point at $\lambda=-1$, thus allowing for the UV completion of the measure sector. This fixed point corresponds to the value obtained by identifying DeWitt's metric from the kinetic term of general relativity, a standard procedure in the literature that otherwise lacks physical motivation. Our results provide such a justification from first principles.
\end{abstract}

\newpage

\section{Introduction}
\label{intro}

Constructing a consistent theory of quantum gravity in the deep UV has been remarkably difficult. This is primarily due to the dual role played by the metric, describing both the gravitational field and the background geometry where matter fields live in. In ordinary quantum field theory, the Minkowski metric is a separate structure fixed from the onset, instrumentally only serving as the background for quantum fields.

The application of quantum field theory to gravity thus usually relies on the split between background and metric fluctuations.
When these fluctuations are well under control, at low energies, the dynamics are fully captured by an effective field theory (EFT) \cite{Donoghue:1994dn}, which is best implemented via the background field method of quantum field theory \cite{Barvinsky:1987uw,Barvinsky:1990up,Barvinsky:1985an}. At Planckian scales, the metric fluctuates wildly and the EFT expansion cannot be truncated. At these energies, quantum gravity could still be described by a quantum field theory should a UV fixed point exist, which would support the asymptotic safety conjecture (see \cite{Niedermaier:2006wt} for a review).

Most modern approaches to quantum gravity as a quantum field theory, be it fundamental or effective, adopts the path integral
\begin{equation}
	\int \d\mu[g] \, e^{i S[g]}
	\label{eq:PI}
\end{equation}
as the definition of the theory, for some classical action $S[g]$. Unlike quantum field theory in flat space, Eq.~\eqref{eq:PI} does not follow from the canonical approach as there is no general way to promote the metric to a quantum operator. 

Even if we leave aside mathematical concerns regarding divergences (which are well-understood under renormalization), questions about the precise definition of the integration measure remain \cite{Fradkin:1973wke,Fujikawa:1983im,DeWitt:2007mi}. This issue exists for any quantum field theory \cite{Kuntz:2024opj,Berganholi:2025hmi}, but only quantum gravity has a non-trivial measure at leading order in power of the fields \cite{DeWitt:2007mi}, as we shall see later. To this date, there is no generally accepted exact form of the gravitational measure.

This old issue has been reignited recently \cite{Donoghue:2020hoh,Branchina:2024xzh,Branchina:2024lai,Branchina:2025kmd,Branchina:2025lqw,Branchina:2025hen,Held:2025vkd,Bonanno:2025xdg}. In Refs.~\cite{Branchina:2024xzh,Branchina:2024lai,Branchina:2025kmd,Branchina:2025lqw,Branchina:2025hen}, the authors make the case for the Fradkin-Vilkovisky measure \cite{Fradkin:1973wke,Kaku:1976xe,Unz:1985wq}, which contain seemingly non-covariant factors $g^{00}$, namely the time-time component of the metric. Following Ref.~\cite{Fradkin:1976xa}, the authors recall that the time-ordering parameter used in the construction of path integrals changes under coordinate transformations. They argue that the $g^{00}$ factor is actually needed to counteract this change in order to keep the diffeomorphism invariance \cite{Branchina:2025lqw}. At the level of the effective action or the equations of motion, these factors must clearly cancel out. Fradkin and Vilkovisky showed that this is indeed the case when the Hessian is correctly treated as a generalized function \cite{Fradkin:1976xa}. On the other hand, Fujikawa proposed a measure that avoids $g^{00}$ entirely, and has been put forward as the true invariant measure \cite{Fujikawa:1979ay,Fujikawa:1983im,Toms:1986sh,Bonanno:2025xdg}.

Geometrically, a measure can be formally defined using the configuration-space metric $G_{ij}$ \cite{Mottola:1995sj,Bern:1990bh,Mazur:1989by,DeWitt:2007mi}:
\begin{equation}
	\d\mu[g] = \mathcal{D}\varphi^i \sqrt{\Det G_{ij}}
	\ .
	\label{eq:measure0}
\end{equation}
At the formal level, this is the direct analog of the volume form encoutered in Riemannian geometry.
One should note that the volume form is not a matter of choice, this measure is the only way of making integrals coordinate-independent in spaces without a symplectic structure. The formal definition \eqref{eq:measure0} clearly does not touch upon the typical issues of infinite-dimensional spaces, thus a covariant regularization is still required. But it serves as a pre-regulated geometrical formulation of path integrals, which is in fact obtained as the continuum limit of Feynman's construction from the operator formalism of non-gravitational field theories (see the Appendix of Ref.~\cite{Kuntz:2024opj} for more details). Futhermore, from Wilson's EFT viewpoint, Eq.~\eqref{eq:PI} should be defined at some cutoff scale, which naturally makes it free from UV divergences.

The problem of quantum gravity then fundamentally depends on the configuration-space metric $G_{ij}$. A quantum theory of gravity is defined not only by the classical action $S[g]$, but by the ordered pair $(S, G)$. DeWitt has introduced the metric \cite{DeWitt:1967yk, DeWitt:2007mi}:
\begin{equation}
	G_{ij}
	=
	\frac{1}{2} (g_{\mu\rho} g_{\nu\sigma} + g_{\mu\sigma} g_{\nu\rho} + \lambda g_{\mu\nu} g_{\rho\sigma})
	\, \delta_g(x,x')
	\ ,
\end{equation}
where we adopted DeWitt's condensed notation, namely $i=(\mu\nu, x)$ and $j=(\rho\sigma, x')$ (see Sec.~\ref{sec:conf} for more details).
The dimensionless parameter $\lambda$, hereby dubbed the DeWitt parameter, is free. For this reason $G_{ij}$ actually forms a one-parameter family of metrics.

The most common practice fixes the configuration-space metric, and hence the DeWitt parameter, from the highest-derivative term of the action \cite{Fradkin:1973wke,Unz:1985wq,Vilkovisky:1984st,Finn:2019aip,Falls:2018olk,Alonso:2015fsp}. For general relativity, for example, this procedure fixes $\lambda=-1$. Although this choice might come across as economical, since it seizes the existence of the action's bilinear form, it lacks physical motivation. Perhaps its best feature is the cancellation of ultralocal divergences \cite{Fradkin:1973wke}, but that can also be achieved by other standard methods. A minority of other works have recognized the ad-hocness of this choice and explored other means of fixing the DeWitt parameter. In Ref.\cite{Odintsov:1991yx}, the role of the DeWitt parameter in determining the radii of spontaneous compactification was analyzed. Ref.\cite{BarberoG:1993cvb} employed its influence on the renormalization group equations to extract physically consistent results, with implications for the decay of the cosmological constant and for dark matter phenomenology. A well-defined effective action in spacetimes with compact extra dimensions was constructed in Ref.~\cite{Huggins:1987zw}, leading to strong constraints on the parameter. More recently, its relevance has been revisited in connection with the conformal factor problem in gravitational path integrals \cite{Liu:2023jvm}.

In this paper, we shall adopt Wilson's EFT to study the renormalization group (RG) flow of the DeWitt parameter $\lambda=\lambda(\Lambda)$ as a function of a covariant cutoff $\Lambda$. We find one UV fixed point at $\lambda=-1$ that corresponds to the general relativistic value obtained from the kinetic term. This shows that the measure sector can be UV completed and justifies the choice $\lambda=-1$ from first principles. In the effective action, the measure correction takes a cosmological-constant like form. At face value, DeWitt's parameter $\lambda$ is found to control such a contribution to the vacuum energy. The observational tiny value for the latter puts the initial value $\lambda_0=\lambda(\Lambda_0)$ in the ballpark of the UV fixed point, corroborating the choice $\lambda=-1$.

This paper is organized as follows. In Sec.~\ref{sec:conf}, we introduce some aspects of the configuration space, particularly regarding DeWitt's notation. We also spell out the different existing conventions in the literature that might cause confusion. In Sec.~\ref{sec:geo}, we review the geometry of the gravitational configuration space and we show how the DeWitt metric follows from an EFT approach at leading order. This also paves the way for the exploration of high-order terms in the configuration-space metric. We then use the DeWitt metric to define the functional measure and we compute its correction to the effective action using the heat kernel regularization in Sec.~\ref{sec:measure}. We then perform the RG analysis of the DeWitt parameter in Sec.~\ref{sec:RG} and draw our conclusions in Sec.~\ref{sec:conc}.

\section{Configuration space: notations and conventions}
\label{sec:conf}

We adopt DeWitt's condensed index notation \cite{DeWitt:2003pm,Parker:2009uva}, in which discrete indices, hereby denoted by capital letters $I$, and spacetime coordinates $x^\mu$ are combined into a single label $i=(x,I)$, which is denoted by the small corresponding letter $i$. Capital-letter indices comprise all types of discrete indices, including gauge, spinor and spacetime ones. For example, for Yang-Mills theories $\varphi^i=A^a_\mu(x)$ and the index $i=(x,\{\mu,a\})$, whereas for gravitational theories $\varphi^i = g^{\mu\nu}(x)$, hence $i=(x,\{\mu,\nu\})$. In general, $\varphi^i=\{\phi(x), A^a_\mu(x), g^{\mu\nu}(x), \ldots\}$ shall denote the whole class of fields present in the model.

Let $\mathcal C$ denote the configuration space over spacetime $\mathcal M$, i.e. the set of all field configurations defined on $\mathcal M$. The configuration space $\mathcal C$ is an infinite-dimensional manifold, whose topology can be heuristically defined by
\begin{equation}
	\mathcal C = \prod_{x\in\mathcal M} \mathcal F(x)
\ ,
\label{toprod}
\end{equation}
where $\mathcal F(x)$ denotes the finite-dimensional target manifold spanned by the set of values $\varphi^i=\varphi^I(x)$ at fixed $x$.
The fields $\varphi^i$ are then treated as coordinates of $\mathcal C$.

\begin{figure}[htb]
	\centering
	\includegraphics[scale=0.5]{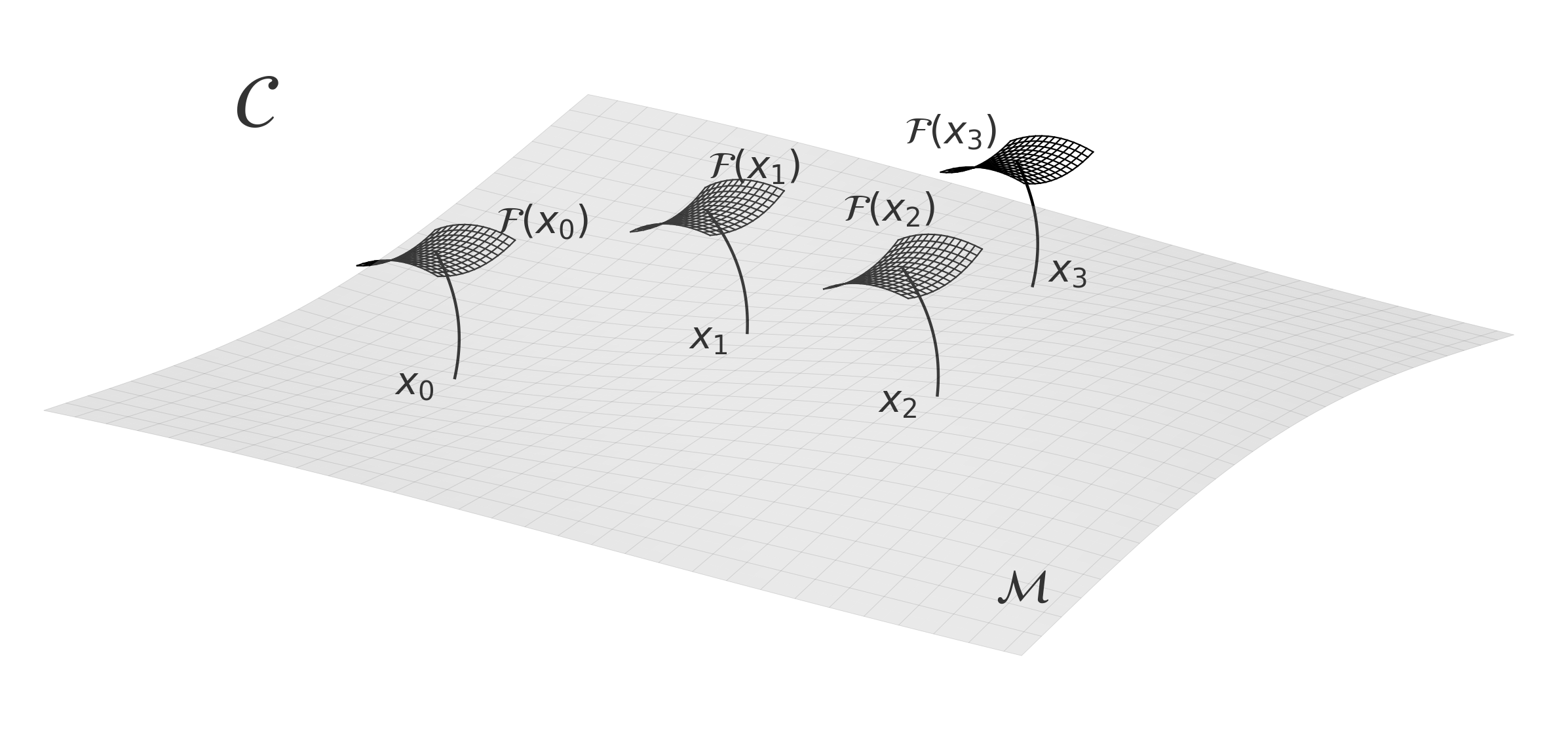}
	\caption{The configuration space $\mathcal C$ is a fiber bundle where the spacetime $\mathcal M$ is the base manifold. All fibers $\mathcal F(x_i)$ are isomorphic to one another.}
\end{figure}

At this point, different conventions come into play in the literature, which we shall outline in the following. First, with the condensed notation, repeated small indices imply sums over discrete indices and integrals over spacetime:
\begin{equation}
	A_i B^i
	=
	\int\d^4x \sqrt{-g} \sum_I A_I(x) B^I(x)
	\ ,
	\qquad
	\text{(geometrical convention)}
	\label{eq:dwsum}
\end{equation}
for arbitrary configuration-space vector $A^i$ and covector $B_i$. Here $A^i, B_i$ are true tensors with their usual transformation property:
\begin{align}
	A'^i &= \frac{\delta\varphi'^i}{\delta\varphi^j} A^j \ ,
	\\
	B'_i &= \frac{\delta\varphi^j}{\delta\varphi'^i} B_j \ ,
\end{align}
under a change of coordinates (field redefinition) $\varphi^i \to \varphi'^i$.
The factor $\sqrt{-g}$ in Eq.~\eqref{eq:dwsum} can be included in the (co)vectors as to form densitized objects:
\begin{align}
	A^i = (-g)^{1/4} A^I(x) \ ,
	\label{do1}
	\\
	B_i = (-g)^{1/4} B_I(x) \ ,
	\label{do2}
\end{align}
thus tensors transform as densities of weight $-1/2$. Eqs.~\eqref{do1} and \eqref{do2} can be readily generalized to any tensor, where each index carries the density $(-g)^{1/4}$.
In this convention, the sumation convention becomes:
\begin{equation}
	A_i B^i
	=
	\int\d^4x \sum_I A_I(x) B^I(x)
	\ .
	\qquad \text{(densitized convention)}
	\label{dc}
\end{equation}

Likewise, the configuration-space line element
\begin{equation}
	\d\mathfrak{s}^2
	=
	G_{ij}\,\d\varphi^i \d\varphi^j
	\label{eq:csmetric}
\end{equation}
can be written as
\begin{equation}
	\d\mathfrak{s}^2
	=
	\int_\Omega\d^nx \sqrt{-g(x)} \int_\Omega\d^nx' \sqrt{-g(x')}
	\,G_{IJ}(x,x')\,\d\varphi^I(x) \d\varphi^J(x')
	\ , \qquad
	\text{(geometrical convention)}
	\label{eq:gcmetric}
\end{equation}
or
\begin{equation}
	\d\mathfrak{s}^2
	=
	\int_\Omega\d^nx \int_\Omega\d^nx'
	\,G_{IJ}(x,x')\,\d\varphi^I(x) \d\varphi^J(x')
	\ ,
	\qquad
	\text{(densitized convention)}
	\label{dcmetric}
\end{equation}
depending on the chosen convention. In the densitized convention, the factors of $\sqrt{-g}$ can absorbed either by the kernel $G_{IJ}(x,x')$, namely:
\begin{equation}
	\sqrt{-g(x)} \sqrt{-g(x')} \, G_{IJ}(x,x')
	\to
	G_{IJ}(x,x')
	\ ,
	\qquad
	\text{(densitized-kernel convention)}
\end{equation}
or by the fields:
\begin{equation}
	\sqrt{-g} \, \d\varphi^i
	\to
	\d \varphi^i
	\ .
	\qquad
	\text{(densitized-field convention)}
\end{equation}
Because the metric $g_{\mu\nu}$ is part of the degrees of freedom, the densitized-field convention involves a non-linear field redefinition, or more formally a change of coordinates in configuration space. In any case, different conventions clearly affect the metric kernel $G_{IJ}(x,x')$, but the full configuration-space metric \eqref{eq:csmetric} remains the same. All geometrical and physical quantities are built from the latter.

The Dirac delta will play a major role in ultralocal metrics, to be introduced later, and it also varies in convention according to its density. We define the covariant Dirac delta by
\begin{equation}
	\int \d^4x' \sqrt{-g(x')} \, \delta_g(x,x') f(x') = f(x)
	\ ,
	\qquad \text{(geometrical convention)}
\end{equation}
which is a biscalar distribution, transforming as a true scalar in both entries $x$ and $x'$.
The standard Dirac delta, defined by
\begin{equation}
	\int \d^4x' \, \delta^{(4)}(x-x') f(x') = f(x)
	\ ,
	\qquad \text{(densitized convention)}
\end{equation}
is rather a bidensity distribution, with weight 0 at $x$ and 1 at $x'$.
It relates to the covariant definition via
\begin{equation}
	\delta_g(x,x') = \frac{\delta^{(4)}(x-x')}{\sqrt{-g(x')}}
	\ .
\end{equation}
Sometimes the symmetric version of the covariant Dirac delta is also adopted:
\begin{equation}
	\delta_s(x,x') = \frac{\delta^{(4)}(x-x')}{(-g(x))^{1/4} (-g(x'))^{1/4}}
	\ ,
\end{equation}
which is a bidensity distribution with weight $1/2$ in each entry.

The coordinate representation $\delta^i_{\ j}$ of the configuration-space identity $\mathbb{1}_{\mathcal C} = \mathbb{1}_{\mathcal F}\otimes \mathbb{1}_x$ depends on the choice of local pairing. For the geometrical convention \eqref{eq:dwsum}, one finds
\begin{equation}
	\delta^i_{\ j}
	=
	\delta^I_{\ J} \, \delta_g(x,x')
	\ ,
	\qquad \text{(geometrical convention)}
\end{equation}
whereas the densitized one, Eq.~\eqref{dc}, gives:
\begin{equation}
	\delta^i_{\ j}
	=
	\delta^I_{\ J} \, \delta^{(4)}(x-x')
	\ .
	\qquad \text{(densitized convention)}
\end{equation}

The configuration-space metric is usually assumed to be ultralocal:
\begin{equation}
	G = \underbrace{\mathbf G}_\text{fiber metric} \otimes \underbrace{\mathbb{1}_x}_\text{continuum direction}
	\ ,
	\label{eq:local}
\end{equation}
which means it takes a block-diagonal form in the continuum direction. The fiber metric $\mathbf G$ acts on the finite-dimensional manifold at fixed $x$.
In coordinates, the metric kernel then becomes:
\begin{align}
	G_{IJ}(x,x')
	&=
	G_{IJ}(\varphi) \, \delta_g(x,x')
	\qquad
	&\text{(geometrical convention)}
	\ ,
	\\
	G_{IJ}(x,x')
	&=
	\sqrt{-g(x)} \, G_{IJ}(\varphi) \, \delta^{(4)}(x-x')
	\qquad
	&\text{(densitized convention)}
	\ ,
\end{align}
and Eqs.~\eqref{eq:gcmetric}--\eqref{dcmetric} collapse into a single integral. We call attention for the subtle notation: the metric kernel $G_{IJ}(x,x')$ and the fiber metric $G_{IJ}(\varphi)$ only differ by their argument.

We shall adopt the geometrical convention throughout this paper, hence all $\sqrt{-g}$ sit in measure. This avoids having to deal with densities and allows us to regard the kernel $G_{ij}$ as the components of a true metric field in configuration space. For convenience, we summarize the different conventions in Table~\ref{tab:conventions}.

\begin{table*}[!htbp]
\setlength{\tabcolsep}{6pt}
\renewcommand{\arraystretch}{1.25}
\footnotesize
\begin{tabularx}{\textwidth}{lYYY}
\toprule
 & \textbf{(A) Tensorial (geometric)} & \textbf{(B) Densitized (Parker--Toms/DeWitt)} \\
\midrule
\textbf{Summation}
&
$\displaystyle A_i B^i = \int \dd^n x\,\sqrt{-g} \; A_I(x)\,B^I(x)$
&
$\displaystyle A_i B^i=\int \dd^n x \, A_I(x) \, B^I(x)$
\\
\textbf{Dirac delta}
&
$\displaystyle \int \dd^n x\,\sqrt{-g(x)}\;\delg(x,x')\,f(x)=f(x')$
&
$\displaystyle \int \dd^n x\,\delta^{(n)}(x-x')\,f(x)=f(x')$
\\
\textbf{Identity kernel}
&
$\delta^i_{\ j} = \delta^I_{\ J} \, \delta_g(x,x')$
&
$\delta^i{}_j = \delta^I{}_J\,\delta^{(n)}(x-x')$
\\
\textbf{Metric kernel}
&
$G_{ij} = G_{IJ}(x,x')$
&
$G_{ij} = \sqrt{-g(x)} \sqrt{-g(x')} \, G_{IJ}(x,x')$
\\
\textbf{Ultralocal metric}
&
$G_{ij} = G_{IJ}(\varphi) \, \delta_g(x,x')$
&
$G_{ij} = \sqrt{-g(x)} \, G_{IJ}(\varphi) \, \delta^{(4)}(x-x')$
\\

\bottomrule
\end{tabularx}
\caption{Comparison of the geometrical and densitized conventions for the summation convention, Dirac delta, the identity components and the metric kernel. In all cases, invariants such as curvatures are convention-independent.}
\label{tab:conventions}
\end{table*}

\section{DeWitt's geometry: a metric over metrics}
\label{sec:geo}

To define the path integral geometrically, in a way that is invariant under field redefinitions, one must endow the configuration space $\mathcal C$ with a geometric structure~\cite{DeWitt:1967yk,Isham:1975ur,Giulini:1993ui,Giulini:2009np,Giulini:1994dx,Giulini:1993ct}. A metric is indeed needed in the definition of a volume form in $\mathcal C$, which is required for the functional integration.
Loop calculations, in particular, involve functional determinants of the Hessian,
\beq
\det \mathcal H_{ij}, \qquad 
\mathcal H_{ij} = \frac{\delta^2 S[\varphi]}{\delta\varphi^i \,\delta\varphi^j} ,
\eeq
which is a bilinear form carrying two covariant indices. The determinant of such an object transforms as a density, hence depends on the chosen basis of the tangent space of $\mathcal C$. Because these indices typically also carry spacetime labels, neither field redefinitions nor coordinate changes leave it invariant. In contrast, determinants of mixed-index operators are basis independent. To obtain invariance one must therefore raise one index of $\mathcal H_{ij}$, and this requires a metric $G_{ij}$ in field space~\footnote{A connection is also required to promote functional derivatives in the Hessian to covariant ones \cite{Vilkovisky:1984st}. Since our focus here is on the measure, which is in fact the dominant contribution (see around Eq.~\eqref{eq:ldhess}), we shall not dwell on this issue.}. In Sec.~\ref{sec:measure}, we shall see that it is precisely the correct measure that brings the Hessian $\mathcal H_{ij}$ into the linear operator form $\mathcal H^i_{\ j} = G^{ik} H_{kj}$.

The configuration-space metric is most commonly identified from the action's kinetic term, whereby one defines the metric as the bilinear form:
\begin{align}
	&\int \mathrm{d}^4x \sqrt{-g} \,
	G_{IJ}(\varphi) \, \partial_\mu\varphi^I \partial^\mu \varphi^J
	\nonumber
	\\
	&=
	\int \mathrm{d}^4x \sqrt{-g(x)} \,
	\int \mathrm{d}^4x' \sqrt{-g(x')} \,
	G_{IJ}(\varphi) \, \delta_g(x,x')\, \frac{\partial \varphi^I(x)}{\partial x^\mu} \frac{\partial \varphi^J(x')}{\partial x'_\mu}
	\ .
	\label{eq:bform}
\end{align}
In this case, the metric would be defined as
\begin{equation}
	G_{ij}
	=
	G_{IJ}(\varphi)
	\, \delta_g(x,x')
	\ .
	\label{ultralocal}
\end{equation}
In this method, ultralocality as defined in Eq.~\eqref{eq:local} follows from the assumption of a local Lagrangian. This definition via the kinetic term can be justified when the phase-space measure
\begin{equation}
	\mathrm{d}\mu[\varphi,\Pi] = \mathcal{M}(\varphi,\Pi) \mathcal{D}\varphi^i \mathcal{D} \Pi_i
\end{equation}
is trivial, i.e. $\mathcal{M}(\varphi,\Pi) \equiv 1$.
In this case, the determinant factor in Eq.~\eqref{eq:measure0} follows upon integration over the canonical momentum $\Pi_i$. However, a trivial phase space (symplectic) measure is too restrictive. Spin systems in quantum mechanics are good examples of theories in which $\mathcal M(\varphi,\Pi)\neq 1$. This argument also relies on the Gaussian integration, thus it cannot be applied to higher-derivative theories where higher powers of $\Pi_i$ is present in the Hamiltonian. For these theories, the path integral is usually defined directly in configuration space and the procedure from Eq.~\eqref{eq:bform} is generalized to the principal symbol (the term with the highest derivative) in the minimal form: $G_{IJ} \Box^n$ for some positive integer $n$. For gauge theories, the minimal form requires a choice of gauge, hence defining the metric from the kinetic term also makes it gauge dependent.

Defining the metric $G_{ij}$ from the kinetic term has its merits. For one, it is an economical choice since the metric is induced by the classical action $S[\varphi]$, which already carries all the physical information at the classical level.  Only the latter is then required to fully determine the quantum theory. However, beyond simplicity, this definition faces limitations for overconstraining the formalism (as we argued above), depending on the gauge choice and lacking physical motivation.

Ref.~\cite{Kuntz:2024opj,Berganholi:2025hmi} introduced a novel approach where the configuration-space metric $G_{ij}$ is constructed using EFT. The fiber metric is then written as an infinite tower of invariant operators:
\begin{equation}
	G_{IJ}(\varphi)
	=
	\sum_{n,m} \frac{c_{n,m}}{\Lambda^n} \, \mathcal O^{(n,m)}_{IJ}(\varphi)
	\ ,
	\label{eq:Gexp}
\end{equation} 
where $c_{n,m}$ are dimensionless free parameters, $\Lambda$ is a cutoff scale, and $\mathcal O^{(n,m)}_{IJ}(\varphi)$ are symmetric rank-2 fiber tensors of mass dimension $n$, built from the fields $\varphi^i$. The index $m$ runs over the different terms with the same mass dimension $n$. The tensors $\mathcal O^{(n,m)}_{IJ}(\varphi)$ are obviously covariant under field redefinitions, but they must be invariant under the underlying symmetries:
\begin{equation}
	\mathcal L_{\xi_A} \, \mathcal O^{(n,m)}_{IJ}
	=
	0
	\ ,
\end{equation}
where $\mathcal L_{\xi_A}$ denotes the Lie derivative with respect to the symmetry group generators $\xi_A$.
From this perspective, the definition of $G_{ij}$ is decoupled from the classical action $S[\varphi]$, but the same symmetry principles of EFT are adopted to define them both independently.
This definition provides a reliable description at energies below the cutoff $\Lambda$, irrespective of the exact functional measure required in a UV-completed theory. In this framework, the free parameters in $G_{ij}$ influence the theory's phenomenology and can be determined through experimental observations rather than solely by their relationship to the kinetic term.

In this EFT approach, $G_{ij}$ is thus regarded as an intrinsic part of the definition of the theory, alongside (but independent of) the classical action.
Different metrics on the configuration space correspond to distinct quantization schemes. 
The complete specification of the theory is therefore given by the pair $(S[\varphi], G)$. Choosing different metrics for the same classical action leads to distinct quantum theories derived from the same classical framework.
Because the metric is no longer tied to the action, ultralocality of the former is no longer a requirement. In fact, if the metric enters the path integral only through the functional measure, its corrections to the quantum theory remain local due to the functional determinant/trace. However, it is still reasonable to assume ultralocality as a natural simplification. This reduces the problem from the infinite-dimensional configuration manifold $\mathcal C$ to the finite-dimensional fiber $\mathcal F$, though infinitely many options for $G_{IJ}$ remain.

For a metric theory of gravity, the basic variables can be taken as either the metric $g_{\mu\nu}$ itself or its inverse $g^{\mu\nu}$. The most common choice in the literature is $\phi^I(x)=g_{\mu\nu}(x)$, hence field-space superindices $I$ correspond to spacetime subindices $(\mu\nu)$ and vice-versa. The choice of the covariant or contravariant metric tensor as the fundamental fields is only conventional. We shall show in Sec.~\ref{sec:measure} that, since they are related by a field redefinition, the covariant path integral (hence physics) remains the same. To avoid the weird index position across field space and spacetime, we shall take $\phi^I(x)=g^{\mu\nu}(x)$.

In the gravitational case, the configuration space is then:
\begin{equation}
	\mathcal C
	=
	\{
	x\mapsto g^{\mu\nu}(x) \,|\, g^{\mu\nu}=g^{\nu\mu}, \, \det g^{\mu\nu} \neq 0 
	\}
	\ .
\end{equation}
At some fixed $x$, the tangent space of the typical fiber $\mathcal F$ is the set of all symmetric rank-2 tensors
\begin{equation}
	T_{g} \mathcal F = {\rm Sym}^2(T_x \mathcal M) = \{h^{\mu\nu} \, | \, h^{\mu\nu} = h^{\nu\mu}\}
	\ ,
\end{equation}
whose dimension is
\begin{equation}
	\dim \Sym(T_x \mathcal M)
	= \frac{n(n+1)}{2}
\end{equation}
in $n$ spacetime dimensions.

To construct the metric $G_{IJ}$ in $\mathcal F$, and hence $G_{ij}$ in $\mathcal C$, we need a bilinear form acting on ${\rm Sym}^2(T_x \mathcal M) \times {\rm Sym}^2(T_x \mathcal M)$. Since $\d\mathfrak{s}^2$ (see Eq.~\eqref{eq:gcmetric}) must be diffeomorphism-invariant, the metric $G_{IJ}$ must transform covariantly. At leading order, Eq.~\eqref{eq:Gexp} yields:
\beq
G_{IJ}
=
c_{0,0} \,
g_{\mu\rho}\, g_{\sigma\nu}
+ c_{0,1} \, g_{\mu\sigma}\, g_{\rho\nu}
+ c_{0,2} \, g_{\mu\nu}\, g_{\rho\sigma}
\ .
\eeq
The symmetry of the metric $G_{IJ}$ in the field-space indices $I=(\mu\nu)$, $J=(\rho\sigma)$, forces $c_{0,0}=c_{0,1}$. This shows that
\begin{align}
	S_{\mu\nu\rho\sigma}
	&\equiv
	g_{\mu\rho}\, g_{\sigma\nu} + g_{\mu\sigma}\, g_{\rho\nu}
	\ ,
	\label{eq:S}
	\\
	B_{\mu\nu\rho\sigma}
	&\equiv
	g_{\mu\nu}\, g_{\rho\sigma}
	\ ,
	\label{eq:B}
\end{align}
form a basis on ${\rm Sym}^2(T_x \mathcal M) \otimes {\rm Sym}^2(T_x \mathcal M)$, namely the space of covariant, parity-even bilinears acting on ${\rm Sym}^2(T_x \mathcal M) \times {\rm Sym}^2(T_x \mathcal M)$.

It is straightforward to figure out that the identity in ${\rm Sym}^2(T_x \mathcal M) \otimes {\rm Sym}^2(T_x \mathcal M)$:
\begin{equation}
	I_{\mu\nu\rho\sigma}
	=
	\frac12 (g_{\mu\rho}\, g_{\sigma\nu} + g_{\mu\sigma}\, g_{\rho\nu})
	\ ,
	\label{eq:id}
\end{equation}
thus $S_{\mu\nu\rho\sigma} = 2 I_{\mu\nu\rho\sigma}$.
Finally, we note that an overall constant in the metric $G_{IJ}$ has no effect on physics because it gets cancelled by the path integral normalization~\footnote{Because the functional determinant itself depends on the spacetime metric, via $\Det A = \exp\int\d^4x \sqrt{-g} \, A$, this cancellation only takes place in the background field method, for $g_{\mu\nu} \to g_{\mu\nu} + h_{\mu\nu}$, where one quantizes the perturbations $h_{\mu\nu}$ around some background $g_{\mu\nu}$. Otherwise, global constant factors in $G_{IJ}$ renormalizes the cosmological constant.}. We then obtain:
\begin{equation}
	G_{IJ}
	=
	\frac12 ( \,
	g_{\mu\rho}\, g_{\sigma\nu}
	+ g_{\mu\sigma}\, g_{\rho\nu}
	+ \lambda \, g_{\mu\nu}\, g_{\rho\sigma} )
	\ ,
	\label{eq:dw}
\end{equation}
where we conveniently chose $c_{0,0}=1/2$ and defined $\lambda = 2 \, c_{0,2}$. Eq.~\eqref{eq:dw} is the original form of the DeWitt metric \cite{DeWitt:1967ub}.

To compute its inverse, we note that Eq.~\eqref{eq:dw} can be written as
\begin{equation}
	G_{IJ}
	=
	I_{\mu\nu\rho\sigma}
	+ \frac{\lambda}{2} B_{\mu\nu\rho\sigma}
	\ .
\end{equation}
Using the operator algebra relations:
\begin{align}
	I^{\mu\nu\rho\sigma} I_{\rho\sigma\alpha\beta} &= I_{\quad \alpha\beta}^{\mu\nu} \ ,
	\\
	I^{\mu\nu\rho\sigma} B_{\rho\sigma\alpha\beta} &= B_{\quad \alpha\beta}^{\mu\nu} \ ,
	\\
	B^{\mu\nu\rho\sigma} B_{\rho\sigma\alpha\beta} &= n \, B_{\quad \alpha\beta}^{\mu\nu} \ ,
\end{align}
which are easily derived from Eqs.~\eqref{eq:S}-\eqref{eq:B}, one finds:
\begin{align}
G^{IJ}
&=
I^{\mu\nu\rho\sigma}
- \frac{\lambda}{2+n\lambda} \, B^{\mu\nu\rho\sigma}
\\
&=
\frac{1}{2}
\left(
g^{\mu\rho}\,g^{\sigma\nu}
+ g^{\mu\sigma}\,g^{\rho\nu}
- \frac{2\lambda}{2+n\lambda}\, g^{\mu\nu}\, g^{\rho\sigma}
\right) ,
\label{eq:minv}
\end{align}
which exists only for $\lambda\neq -2/n$. As we shall see in the next section, fixing $G_{IJ}$ from the kinetic term gives the same structure \eqref{eq:dw} with $\lambda_{GR}=-1$ for general relavity. In fact, because the construction of Eq.~\eqref{eq:dw} is based purely on symmetry principles, the metric identification from the action gives exactly \eqref{eq:dw} for any diffeomorphism-invariant model, with $\lambda$ parameterizing the theory. We shall however leave $\lambda$ free to be bounded phenomenologically.  

\section{The gravitational functional measure}
\label{sec:measure}

Path integrals have become a standard tool across quantum field theory, appearing in contexts from condensed matter to particle physics. Their usefulness is particularly evident in gauge theories, where they maintain manifest Lorentz covariance and gauge invariance. Within Wilson's framework of EFT, path integrals also provide the natural formulation of quantum field theory itself, which is finite from the onset because of the existence of an intrinsic cutoff scale. In this paper, we shall adopt Wilson's approach for renormalization, tracking the running of the measure with the cutoff.

In Wilson's viewpoint the theory is defined only below some physical threshold $\Lambda_\text{phys}$, and the construction introduces a finite cutoff $\Lambda < \Lambda_\text{phys}$. We shall take $\Lambda_\text{phys} \sim M_p$ as the Planck mass. The generating functional reads
\begin{equation}
	Z_\Lambda[J]
	=
	\int^\Lambda \mathrm{d}\mu[\varphi] 
	\, \exp\!\left[i\Big(S[\varphi]+J_i\varphi^i\Big)\right],
	\label{Z}
\end{equation}
where $S[\varphi]$ is the classical action and $J_i$ denotes external sources. The scale $\Lambda_\text{phys}$ represents an intrinsic short-distance regulator, e.g.~the lattice spacing in condensed matter or the string scale in high-energy theory. Because of $\Lambda_\text{phys}$, ultraviolet divergences never arise, and renormalization simply enforces the $\Lambda$-independence of observables:
\begin{equation}
	\Lambda \frac{d Z_\Lambda[J]}{d \Lambda} = 0
	\ .
	\label{RG}
\end{equation}

Even though the Wilsonian construction avoids UV infinities, the integration measure $\mathrm{d}\mu[\varphi]$ remains poorly specified. Most treatments assume it factorizes trivially,  
\begin{equation}
	\mathcal{D}\varphi^i
	=
	\prod_i \mathrm{d}\varphi^i
	\ ,
	\label{trivialm}
\end{equation}
but in fact the measure is sensitive to the geometry and topology of configuration space. Singular field configurations or nontrivial gauge constraints typically obstruct the naive factorized form \eqref{trivialm}, since the field space is neither simply connected nor flat in general. These features are already anticipated from the canonical phase-space structure (see Appendix of Ref.~\cite{Kuntz:2024opj}).  

Geometrically, the measure is defined via the metric~\footnote{We shall denote by capital initial letters, such as $\Det$ and $\Tr$, functional operations on configuration-space objects and by small initial letters, such as $\det$ and $\tr$, operations on the finite-dimensional bundle.}:
\begin{equation}
    \mathrm{d}\mu[\varphi] = \mathcal{D}\varphi^i \,\sqrt{\Det G_{ij}}
    \ .
    \label{measure}
\end{equation}
We recall that, in the standard finite-dimensional manifolds, the volume form $\omega = \d^4x \sqrt{-g}$ is not a matter of choice. Rather, it is the sole way to consistently define integrations geometrically (i.e. coordinate-independent), irrespective of the manifold being curved. The same happens here, with Eq.~\eqref{measure} being required for coordinate independence in $\mathcal C$. In physical terms, this is just the statement of independence on the field parameterization $\varphi^i$, thereby turning field redefinitions into mere change of coordinates in $\mathcal C$.

The covariant definition \eqref{measure} has plenty of interesting features. For one, it solves the ambiguity in the choice of basic gravitational field among $g_{\mu\nu}$, $g^{\mu\nu}$ or even the densitized variables $\hat g_{\mu\nu} \equiv (-g)^r g_{\mu\nu}$, $\hat g^{\mu\nu} \equiv (-g)^{-r} g^{\mu\nu}$. Although it is clear that classical physics does not depend on this choice, quantum mechanically one would be integrating over a field with different transformation properties, thus defining different quantum theories \cite{Ohta:2016npm,Ohta:2016jvw}. These different parameterizations of the gravitational degrees of freedom are connected by field redefinitions:
\begin{align}
	g_{\mu\nu} &\to g^{\rho\sigma} = g^{\rho\mu} g^{\sigma\nu} g_{\mu\nu} \ ,
	\\
	g_{\mu\nu} &\to \hat g_{\mu\nu} = (-g)^r g_{\mu\nu} \ ,
\end{align}
and so on. Therefore, any possible Jacobian from $\mathcal D g_{\mu\nu}$ gets cancelled by the inverse Jacobian from $\sqrt{\Det G_{ij}}$.

Although expression \eqref{measure} is generally nontrivial, it is a common practice to appeal to dimensional regularization and effectively reduce it to unity. The argument proceeds from
\begin{align}
    \Det G_{ij}
    &=
    e^{\Tr[\log G_{ij}]}
    \\
    &=
    \exp\left(\delta^{(n)}(0)\int\mathrm{d}^nx \,\log\det G_{IJ}\right),
    \label{dimr}
\end{align}
valid for ultralocal metrics of the form \eqref{ultralocal}. Since scaleless integrals vanish under dimensional regularization, one often sets $\delta^{(n)}(0)=0$ and concludes $\Det G_{ij}=1$. This step, however, is problematic: Eq.~\eqref{dimr} involves products of $\delta^{(n)}(0)=0$ and divergent terms from $\log\det G_{IJ}$. A careful treatment shows that $\Det G_{ij}\neq 1$ in general when the UV limit is properly taken with covariant regulators. Anomalies serve as the most common example of this fact \cite{Fujikawa:1979ay}.

This can indeed be seen by using a covariant cutoff $\Lambda$. We shall here adopt the proper time regularization with $s=\Lambda^{-2}$, so that:
\begin{align}
	\Tr[\log \hat G]
	&\to
	\Tr[(\log \hat G) e^{i s (\hat H+i\epsilon)}]
	\\
	&=
	\int \d^4x \sqrt{-g}
	\tr[\log (\mathbf G) \, \mathbf K(x,x; s)]
	\ ,
\end{align}
for some diffeomorphism-invariant operator $\hat H$ acting on the space of metric fluctuations $\Sym(T \mathcal M)$. We have used hats, e.g. $\hat O = (O_{ij})$, to abbreviate abstract operators acting on fields and bold, e.g. $\mathbf O = (O_{IJ})$, to abbreviate fiber matrices acting on tensors. The heat kernel in the coincident limit is given by the Schwinger-DeWitt expansion \cite{DeWitt:2003pm,Vassilevich:2003xt}:
\begin{equation}
	\mathbf K(x,x;s) = \frac{-i}{(4\pi s)^{2}}\sum_{n=0}^{\infty} (is)^{n} \mathbf a_{n}(x; \hat H)
	\ ,
\end{equation}
where $\mathbf a_n(x; \hat H)$ denote the heat kernel coefficients of the operator $\hat H$.
The functional determinant then reads:
\begin{equation}
	\Det \hat G(\Lambda)
	=
	\exp{
		\int\mathrm{d}^4x
		\tr[
			\log \mathbf G \, \frac{- i \Lambda^4}{(4\pi)^2} \left(\mathbf a_0 + i \Lambda^{-2} \mathbf a_1 - \Lambda^{-4} \mathbf a_2 + \cdots \right)
		]
	}
	\ ,
	\label{eq:detreg}
\end{equation}
where the ellipsis regard negative powers of $\Lambda$. In ordinary renormalization, power divergences (here proportional to $\mathbf a_0$ and $\mathbf a_1$) are removed by counterterms, but a finite term, due to $\mathbf a_2$, remains. Thus, after renormalization:
\begin{equation}
	\Det G_{ij} \Big|_{\rm finite}
	=
	\exp{
		\frac{i}{(4\pi)^2}
		\int\mathrm{d}^4x \sqrt{-g}
		\tr[
			\log (\mathbf G) \, \mathbf a_2
		]
	}
	\ ,
	\label{eq:finite}
\end{equation}
which shows that the measure cannot be trivialized by the misuse of dimensional regularization.

Since we shall adopt Wilson's EFT, where one keeps track of the dependence on $\Lambda$ rather than subtracting it, all terms in the heat kernel expansion plays a role, being now distinguished between relevant, irrelevant and marginal corrections. Recalling that $\mathbf a_n\sim \mathcal{R}^{2n}$, where $\mathcal R$ denotes the generalized curvature (including spacetime curvature, gauge field strength and scalar potentials), beyond $\mathbf a_0 = \mathbb 1$ all higher-order terms are suppressed by the cutoff $\Lambda \lesssim M_p$. We shall thus keep only the leading order (and relevant) term in Eq.~\eqref{eq:detreg}.

Exponentiating $\sqrt{\Det G_{ij}}$ in \eqref{measure}, allows one to write:
\begin{equation}
	\int \d\mu[g] \, e^{\frac{i}{\hbar} S[g]}
	=
	\int \mathcal D g_{\mu\nu} \, e^{\frac{i}{\hbar} \left( S[g] - \frac{i \hbar}{2} \log \Det G_{ij} \right)}
	\ ,
	\label{eq:pi1}
\end{equation}
where we have reinstated $\hbar$ momentarialy just to show that the measure correction is one-loop exact. To compute the path integral in Eq.~\eqref{eq:pi1}, we work in the background field formalism
\begin{equation}
	g_{\mu\nu}
	\to
	g_{\mu\nu}
	+ h_{\mu\nu}
	\ ,
\end{equation}
and impose a gauge condition $F_\mu[g,h]=0$. This introduces the gauge-fixing functional
\begin{equation}
	S_{gf}
	=
	\frac{M_p^2}{2}\int d^4x\,\sqrt{-g}\;\frac{1}{2\alpha}\,F_\mu F^\mu
	\ ,
\end{equation}
for the gauge parameter $\alpha$.
The variation of $F_\mu$ under an infinitesimal gauge transformation produces the Faddeev--Popov operator, whose determinant is represented by the ghost action $S_{gh}[c,\bar c]$. Expanding the gauge-fixed action
\begin{equation}
	S_{\rm full}
	=
	S
	+ S_{gf}
	+ S_{gh}
\end{equation}
to second order in the fluctuations yields the Hessians $\mathcal{H}_{ij}=S_{{\rm full},ij}$. Because the path integral becomes Gaussian at this order, integrating out the fluctuations and ghosts gives the one-loop result:
\begin{align}
	\Gamma[\varphi^i]
	&=
	S[\varphi^i]
	- \frac{i}{2} \log\Det G_{ij}
	+ \frac{i}{2} \log\Det \mathcal{H}_{ij}
	\label{inv1loop}
	\\
	&=
	S[\varphi^i]
	+ \frac{i}{2} \log\Det \mathcal{H}^i_{\ j}
	\ .
	\label{invact}
\end{align}
The Hessian enters only through the combination $\mathcal{H}^i_{\ j} = G^{ik} \mathcal{H}_{kj}$ thanks to $G_{ij}$, turning the determinant of a bilinear form into that of a linear operator~\cite{Toms:1986sh,Ellicott:1987ir}.

In the following, we shall compute $\Det (G^{ik} \mathcal{H}_{kj})$ using the DeWitt metric, which amounts on using the regularization \eqref{eq:detreg} for the operator $\hat G^{-1} \hat{\mathcal H}$ in place of $\hat G$. For pure gravity, the classical dynamics follows from the Einstein--Hilbert action
\begin{equation}
	S = \int\mathrm{d}^4x \sqrt{-g}\,\frac{M_p^2}{2} R
	\ ,
	\label{eq:EH}
\end{equation}
with Planck mass $M_p$ and Ricci scalar $R$. We pick the de Donder gauge condition:
\begin{equation}
	F_\mu = \nabla^\nu h_{\mu\nu} - \tfrac12\,\nabla_\mu h
	\ .
\end{equation}
For the Feynman--'t Hooft gauge, i.e. $\alpha=1$, one finds:
\begin{equation}
	S^{(2)}+S_{gf}
	= \frac{M_p^2}{8}\int d^4x\,\sqrt{-g}\;
	h^{\mu\nu}\,\mathcal{H}_{\mu\nu}{}^{\rho\sigma}\,h_{\rho\sigma}
	\ ,
\end{equation}
in which case the Hessian takes a minimal form:
\begin{align}
\mathcal H_{\mu\nu,\rho\sigma}
&=
- \Box
O_{\mu\nu,\rho\sigma}
\;+
U_{\mu\nu,\rho\sigma}
\ ,
\label{eq:hess}
\end{align}
with
\begin{align}
	\label{eq:K}
	O_{\mu\nu,\rho\sigma}
	&=
	\frac12 \! \left(
		g_{\mu\rho} g_{\nu\sigma}+ g_{\mu\sigma} g_{\nu\rho}
		- g_{\mu\nu} g_{\rho\sigma}
	\right)
	\ ,
	\\
	U^{\mu\nu}_{\ \ \rho\sigma}
	&=
	2 R \, O^{\mu\nu}_{\ \ \rho\sigma}
	+ g^{\mu\nu} R_{\rho\sigma} + g_{\rho\sigma} R^{\mu\nu}
	- 2 \delta^{(\mu}_{(\rho} R^{\nu)}_{\sigma)}
	- 2 R^{(\mu \quad \nu)}_{\quad (\rho \quad \sigma)}
	\ .
\end{align}

To put the Hessian in canonical form, we factor $\mathbf O$ out in Eq.~\eqref{eq:hess}, namely:
\begin{equation}
	\log\det(\mathbf G^{-1} \mathcal H)
	=
	\log\det(\mathbf G^{-1} \mathbf O)
	+ \log\det(-\Box + \mathbf O^{-1} \mathbf U)
	\ .
	\label{eq:ldhess}
\end{equation}
The last term of \eqref{eq:ldhess}, along with the ghost contribution from $(\mathcal H^{gh})^i_{\ j}$, is the typical one-loop contribution. They yield the one-loop divergences, via the Schwinger-DeWitt expansion \cite{DeWitt:2003pm}, and the non-local form factors, by the Barvinsky-Vilkovisky covariant perturbation theory~\cite{Barvinsky:1987uw,Barvinsky:1985an}. Both expansions produce higher curvature terms, which are suppressed at low energies. The dominant contribution arises from the first term of \eqref{eq:ldhess}, which is derivative-free and reflects the non-trivial functional measure. We shall thus focus on this contribution.

From Eqs.~\eqref{eq:minv} and \eqref{eq:K}, one finds:
\begin{equation}
	\mathbf G^{-1} \mathbf O
	=
	\mathbb{1}
	+ \frac12 \left(\frac{\lambda}{2\lambda + 1} - 1 \right) \mathbf B.
	\label{eq:GK}
\end{equation}
We note that factors of $\sqrt{-g}$, when present (see Sec.~\ref{sec:conf}), cancel out in Eq.~\eqref{eq:GK} because $\mathbf G^{-1}\propto 1/\sqrt{-g}$ and $\mathbf O\propto \sqrt{-g}$.
Introducing the projectors onto the trace and traceless sectors:
\begin{align}
	\mathbf P_T
	&= \frac14 \mathbf B
	\ ,
	\\
	\mathbf P_{TL}
	&= \mathbb{1} - \frac14 \mathbf B
	\ ,
\end{align}
Eq.~\eqref{eq:GK} can be written as
\begin{equation}
	\mathbf G^{-1} \mathbf O
	=
	\mathbf P_{TL}
	- \frac{1}{2\lambda + 1} \mathbf P_T
	\ .
\end{equation}
Since a symmetric rank-2 tensor in four dimensions decomposes into a traceless part (9 components) and a trace part (1 component), the eigenvalues of $\mathbf G^{-1} \mathbf O$ are
\begin{align}
	\lambda_{TL}
	&=
	1 &\text{(multiplicity 9)} \ ,
	\\
	\lambda_{T}
	&=
	- \frac{1}{2\lambda + 1} &\text{(multiplicity 1)}
	\ .
\end{align}
The determinant then reads:
\begin{equation}
	\det (\mathbf G^{-1} \mathbf O)
	=
	- \frac{1}{2\lambda + 1}
	\ .
	\label{eq:detGO}
\end{equation}
Using the regularization \eqref{eq:detreg} for $\hat G^{-1} \hat O$ at leading order $\Lambda^4$, together with Eqs.~\eqref{invact} and \eqref{eq:detGO}, gives:
\begin{equation}
	\Gamma_{\rm eff}[g]
	=
	\int\mathrm{d}^4x \sqrt{-g}\,
	\left[
		\frac{M_p^2}{2} R
		+ \Lambda_m
	\right],
	\label{1piac2}
\end{equation}
and the measure correction yields a cosmological-constant like term in the effective action given by:
\begin{equation}
	\Lambda_m
	=
	\frac{\Lambda^4}{2(4 \pi)^{2}}
		\log \frac{-1}{2\lambda + 1}
	\ .
	\label{eq:cc}
\end{equation}
The eigenvalue in the trace direction can be negative depending on the value of $\lambda$, thus making the log in Eq.~\eqref{eq:cc} generally complex:
\begin{equation}
	\log \frac{-1}{2\lambda + 1}
	=
	\log \frac{1}{|2\lambda + 1|}
			+ i \pi \theta(2\lambda+1)
	\ .
\end{equation}
Because this imaginary term has no momentum/derivative dependence, it does not satisfy Cutkosky rules, therefore its presence would render the theory non-unitary. The physical region of the parameter space is thus $2\lambda+1<0$. In Sec.~\ref{sec:RG}, we will show that $\mathrm{sgn(2\lambda+1)}$ is invariant along the RG flow, hence if the theory is unitary for some initial cutoff $\Lambda_0$, it remains so for any $\Lambda$.

\section{RG analysis}
\label{sec:RG}

\begin{figure}[h!]
	\centering
	\includegraphics[scale=0.5]{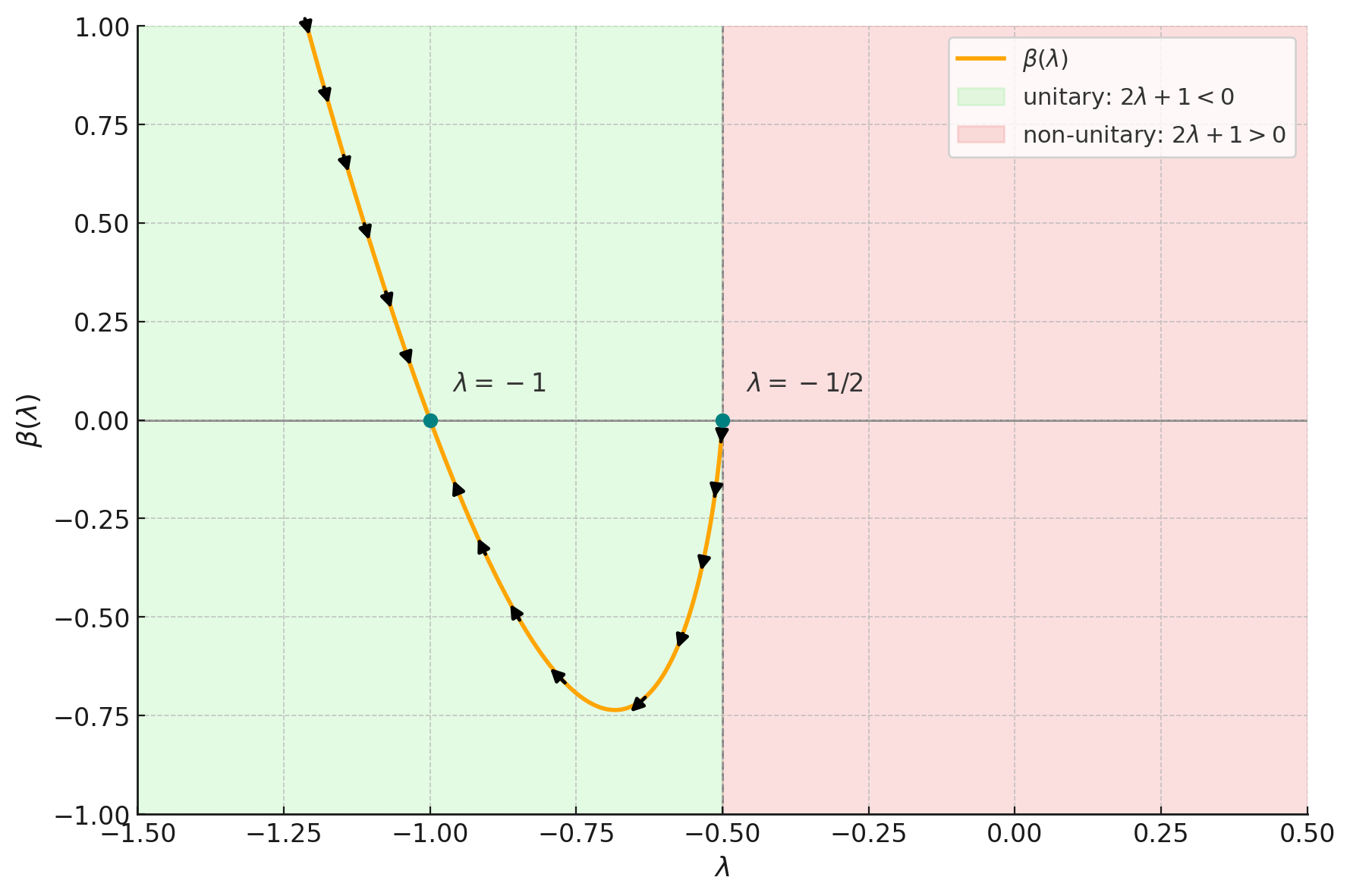}
	\caption{Phase portrait for the RG equation \eqref{RGa} with the beta function $\beta_\lambda$ (orange curve) given by Eq.~\eqref{eq:beta}. The fixed points are marked with green blobs, with the arrows (pointed towards the UV) showing their (in)stability. The figure also shows the unitary (in green) and the non-unitary (in red) regions of parameter space. In the non-unitary region, $\beta_\lambda$ is complex and the orange curve illustrates its real part.}
	\label{fig:portrait}
\end{figure}

In our truncation, the RG invariance in Eq.~\eqref{RG} translates in the the RG invariance of Eq.~\eqref{eq:cc}:
\begin{equation}
	\Lambda \frac{\d \Lambda_m}{\d \Lambda} = 0
	\ .
	\label{RGm}
\end{equation}
This promotes the DeWitt parameter $\lambda=\lambda(\Lambda)$ to a running coupling:
\begin{equation}
	\Lambda \frac{\d \lambda}{\d\Lambda}
	=
	\beta_\lambda
	\ ,
	\label{RGa}
\end{equation}
for the beta function (see the phase portrait in Fig.~\ref{fig:portrait}):
\begin{equation}
	\beta_\lambda
	= -2 (2\lambda + 1) \log |2\lambda + 1|
	\ .
	\label{eq:beta}
\end{equation}
We note that the volume term in Eq.~\eqref{eq:detreg} is entirely fixed by the coefficient $\mathbf a_0$, since the higher coefficients $\mathbf a_{n \ge 1}$ multiply curvature-dependent invariants such as $R$, $R^2$, etc. As a consequence, the $\lambda$-dependence of the measure-induced contribution to the cosmological term, and hence the beta function \eqref{eq:beta} defined via this sector, is determined solely by $\mathbf a_0$. Including higher-order terms in Eq.~\eqref{eq:detreg} modifies the running of curvature-dependent couplings, but does not change Eq.~\eqref{eq:beta}.

From Eq.~\eqref{eq:beta} one can readily identify the fixed points:
\begin{align}
	\lambda_{\rm UV} &= -1 \ ,
	\\
	\lambda_{\rm IR} &= -\frac12 \ .
\end{align}
Their stability follows from the linearized RG equation:
\begin{equation}
	\frac{\mathrm{d}\beta_\lambda}{\mathrm{d}\lambda}
	= -4 \left( \log|2\lambda + 1| + 1 \right)
	\ ,
\end{equation}
which shows that $\lambda_{\rm UV}$ is UV attractive ($\beta_\lambda'(\lambda_{\rm UV}) < 0$), whereas $\lambda_{\rm IR}$ is singular and UV repulsive ($\beta_\lambda'(\lambda_{\rm IR}) > 0$) from both sides. Amusingly, both fixed points correspond to important values for DeWitt metric: $\lambda_{\rm UV}$ matches Einstein-Hilbert's kinetic term and $\lambda_{\rm IR}$ makes the DeWitt metric degenerate (non-invertible).

\begin{figure}[h!]
	\includegraphics[scale=0.5]{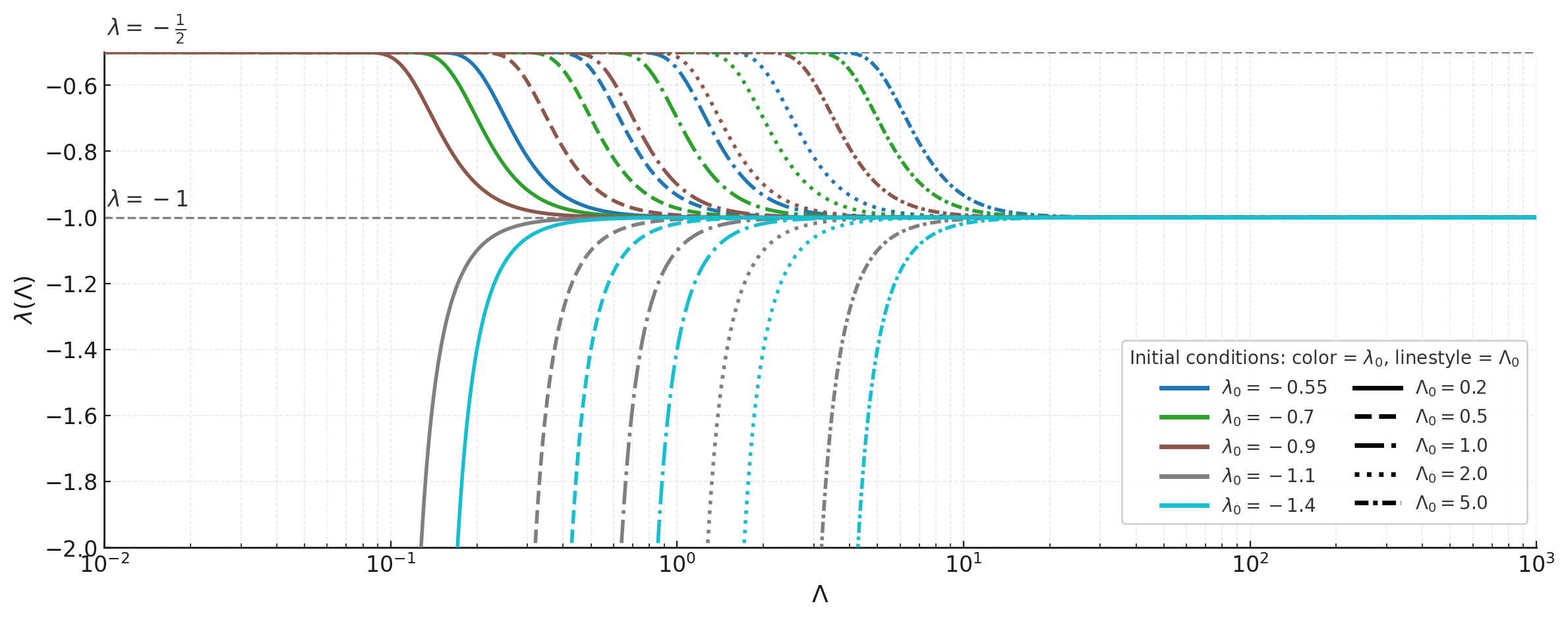}
	\caption{RG trajectories given by the solution \eqref{eq:RGsol} for different choices of initial data $(\Lambda_0,\lambda_0)$. Different colors correspond to different choices of $\lambda_0$, whereas different linestyles correspond to different choices of $\Lambda_0$. All trajectories asymptotically flow to the UV fixed point at $\lambda=-1$.}
	\label{fig:sol}
\end{figure}

Since the RGE~\eqref{RGa} is one-dimensional and $\Lambda_m$ is conserved along RG trajectories, the system is integrable and can be solved exactly. One finds (see Fig.~\ref{fig:sol}):
\begin{equation}
	2\lambda(\Lambda)+1=-\Big[-\big(2\lambda_0+1\big)\Big]^{(\Lambda_0/\Lambda)^4}
	\ ,
	\label{eq:RGsol}
\end{equation}
for some initial cutoff $\Lambda_0 > \Lambda$ and $\lambda_0 = \lambda(\Lambda_0)$, or in terms of the RG invariant $\Lambda_m$:
\begin{equation}
	2\lambda(\Lambda)+1=-\,e^{-2(4\pi)^2\Lambda_m/\Lambda^4}
	\ .
	\label{arun}
\end{equation}
Eqs.~\eqref{eq:RGsol}-\eqref{arun} show that $\mathrm{sgn(2\lambda+1)}$ depends only on the initial condition, thus remaining constant (and negative) along the RG evolution.
Hence, $\lambda(\Lambda)$ does not cross the singular value $\lambda_{\rm IR} = -\frac12$, only approaching it asymptoticaly for $\Lambda\to 0$. This creates two independent regions, corresponding to the unitary and non-unitary regimes (Fig.~\ref{fig:portrait}), with the UV fixed point lying on the unitary side. An initial (non-)unitary system remains so along the entire RG flow. For $\lambda_0 > \lambda_{\rm IR}$, the theory is then unphysical. On the other hand, all trajectories starting off from some $\lambda_0 < \lambda_{\rm IR}$ will approach the fixed point $\lambda_{\rm UV}$ in the UV (Fig.~\ref{fig:sol}). This happens because the RG equation \eqref{RGa} is one-dimensional, thus the basin of attraction of $\lambda_{\rm UV}$ spans the entire unitary domain $\lambda<-\frac12$. Moreover, if the measure correction in Eq.~\eqref{1piac2} were to be interpreted as vacuum energy, its positivity would constrain $\lambda$ to
\begin{equation}
	-1 \leq \lambda \leq -\frac12
	\ ,
\end{equation}
with the fixed points lying on the boundaries.

For consistency within the Wilsonian RG program, the scales of interest must be well below the cutoff, thus $\Lambda_m^{1/4} \ll \Lambda$. Eq.~\eqref{arun} then shows that the DeWitt parameter is exponentially close to the UV fixed point:
\begin{equation}
	\lambda(\Lambda)
	\approx
	- 1
	\ ,
	\label{eq:asym}
\end{equation}
where higher-order terms $\mathcal O\left(\tfrac{\Lambda_m}{\Lambda^4}\right)$ are suppressed by inverse powers of the cutoff, hence the approximate sign.
Should we take, for example, the cutoff as the Planck mass $\Lambda = M_p \sim 10^{19} \, {\rm GeV}$, we would find $\Lambda_m/\Lambda^4 \sim 10^{-70}$. Any reasonable sub-Planckian value of $\Lambda_m$ puts $\lambda$ in the immediate ballpark of the UV fixed points. In particular, in the continuum limit \eqref{eq:asym} becomes exact: $\lambda$ lies precisely at one of the fixed points.

The RG thus dynamically selects out $\lambda=-1$ as the unique value for the DeWitt parameter at the UV aymptotic regime. Like we mentioned before, this value matches the value one obtains by defining the configuration-space metric from Einstein-Hilbert's quadratic form. Only this time $\lambda=-1$ is not an artificial choice, but a robust prediction of the RG evolution. No fine-tuning is required.

This also shows that the configuration space is curved. The flat configuration-space metric corresponds to $\lambda=0$, which aligns to the usual approach where the measure \eqref{measure} in cartesian coordinates is the trivial one:
\begin{equation}
    \mathrm{d}\mu[g] = \mathcal{D} g_{\mu\nu}
    \ .
\end{equation}
But the value $\lambda=0$ lies on the non-unitary side of the evolution (see Fig.~\ref{fig:portrait}), thus preventing the definition of the measure in this naive way.

\section{Conclusions}
\label{sec:conc}

Albeit largely ignored, the functional measure is of utmost importance in the quantum theory, be it gravitational or otherwise. The measure is in fact part of the definition of the path integral, so fully understanding it might be they key to address open problems and unravel new physics. While its definite form remains unknown, EFT paves the way for a model-independent approach that relies solely on symmetries.

We showed that at the DeWitt truncation, the RG contains IR and UV fixed points, namely $\lambda_{\rm UV}=-1$ and $\lambda_{\rm IR}=-1/2$. The former allows for the UV completion of the measure sector, proving that the DeWitt truncation could actually be defined in the continuum limit $\Lambda\to\infty$. At finite cutoff, unitarity restricts the parameter space to $\lambda<-1/2$, thus excluding the flat configuration-space, i.e. $\lambda=0$, on physical grounds. The value $\lambda_{\rm UV}=-1$ of this UV fixed point is precisely the one obtained by defining the configuration-space metric from the kinetic term of general relativity.

 This procedure of identifying the metric from the kinetic term, albeit standard in the literature, lacks physical motivation and proper justification. Our result provides such a justification by showing that $\lambda_{\rm UV}=-1$ is the only UV fixed point of the DeWitt parameter. For any unitary initial data, all trajectories flow asymptotically to $\lambda=\lambda_{\rm UV}$. The measure remains well-defined for arbitrarily high energies without violating unitarity. Should we interpret the measure correction as vacuum energy, experimental data puts stringent bounds on the correction, pointing at an almost vanishing $\Lambda_m \sim 0$ and corroborating that initial RG data must land exponentially close to the UV fixed point.

We note that higher order terms to the configuration-space metric most likely change the above scenario because of the departure from the DeWitt metric. This could shift the fixed points, create new ones or remove them. Possible extensions to higher orders could allow for cross-terms between gravitational and matter fields, derivative terms (including curvatures) and deviations from ultralocality. Moreover, our justification for defining the metric from the action does not automatically extend to other theories, in which a case-by-case study is needed. The details of these possibilities shall be left for future investigations.

\section*{Acknowledgments}
IK is grateful to the National Council for Scientific and Technological Development -- CNPq (Grant Nos. 303283/2022-0, 401567/2023-0 and 200564/2025-0) for partial financial support. The author thanks Stefano Liberati for many useful discussions and for comments on earlier versions of this work.

%
%
%
%

%
\end{document}